\documentclass[letter,twocolumn]{jpsj3}
\usepackage{txfonts,braket,amsmath,bm,here,color}

\everymath{\displaystyle}

\title{Rotational Response Induced by Electric Toroidal Dipole}

\author{Akimitsu Kirikoshi and Satoru Hayami}
\inst{Department of Physics, Hokkaido University, Sapporo 060-0810, Japan}

\abst{
  A ferroaxial ordering, which appears without mirror symmetry parallel to an electric axial moment, is described by a ferroic alignment of the electric toroidal (ET) dipole rather than the conventional electric and magnetic dipoles.
  Although its emergence requires neither spatial inversion nor time-reversal symmetry breakings, unconventional transverse responses between the conjugate physical quantities have been proposed, which are qualitatively different from those in multiferroic systems without both spatial inversion and time-reversal symmetries.
  We theoretically investigate a general relationship between ferroaxial ordering and its characteristic response tensor. 
  We show that various rotational responses corresponding to an antisymmetric tensor component are related to the ferroaxial ordering based on symmetry analysis. 
  Among them, we propose that second-order nonlinear magnetostriction, where the strain is induced by a second-order magnetic field, is one of the experimental setups to identify the ferroaxial ordering. 
  We show its temperature and magnetic-field-angle dependence by analyzing a fundamental $d$-orbital model under the tetragonal symmetry. 
}

\begin{document}
\maketitle

{\it{Introduction.}}---
Electronic orderings by spontaneous symmetry breakings through electron correlation have long been studied in condensed matter physics.
Among them, a ferroic order parameter, as exemplified by ferromagnetic and ferroelectric ones, has drawn the attention of researchers in various fields since it leads to not only understanding of fundamental physical properties but also applications to electronic/spintronic devices.
One of the recent examples is the ferrotoroidal ordering in the absence of both spatial inversion ($\mathcal{P}$) and time-reversal ($\mathcal{T}$) symmetries, which becomes a source of a linear magneto-electric effect, where the electric (magnetic) field induces magnetization (electric polarization)~\cite{Spaldin_2008}.
These findings bring about new properties, which opens a further rich possibility of exploring functional materials.

More recently, a new-type ferroic order parameter has been proposed, referred to as ``ferroaxial'' (or ``ferro-rotational'')~\cite{natphys.16.42}.
The ferroaxial ordering corresponds to the rotational distortion of the atomic arrangement in the crystal from the symmetry viewpoint, whereas $\mathcal{P}$ and $\mathcal{T}$ symmetries are not necessary broken.
This ordering has been observed in some materials, e.g., $\mathrm{Cu_{3}Nb_{2}O_{8}}$~\cite{PhysRevLett.107.137205}, $\mathrm{CaMn_{7}O_{12}}$~\cite{PhysRevLett.108.067201}, $\mathrm{RbFe(MoO_{4})_{2}}$~\cite{natphys.16.42, PhysRevMaterials.5.124409}, $\mathrm{NiTiO_{3}}$~\cite{PhysRevMaterials.5.124409,natcommun.11.4582, npjqm.7.106, PhysRevB.107.L180102, pnas.2303251120}, $\mathrm{Ca_{5}Ir_{3}O_{12}}$~\cite{JPSJ.89.054602, JPSJ.90.063702, JPSJ.92.033702, JPSJ.92.063601}, $\mathrm{BaCoSiO_{4}}$~\cite{BaCoSiO4}, $\mathrm{K_{2}Zr(PO_{4})_{2}}$~\cite{acs.chemmater.2c03540}, $\mathrm{Na_{2}Hf(BO_{3})_{2}}$~\cite{acs.chemmater.3c00624}, and Na-superionic conductors~\cite{jacs.3c00797}.
Meanwhile, the external field that directly controls the order parameter of the ferroaxial order has not been fully elucidated owing to the lack of conjugate electromagnetic fields.

In parallel with the experimental findings of the ferroaxial ordering, its atomic-scale electronic order parameter has been clarified based on the multipole representation~\cite{PhysRevB.98.165110}: the electric toroidal (ET) dipole defined by the vector product of orbital and spin-angular momentum operators~\cite{JPSJ.89.104704}.
Furthermore, various off-diagonal responses of the conjugate physical quantities have been proposed based on symmetry and model analyses, e.g., intrinsic spin-current generation~\cite{JPSJ.91.113702, PhysRevMaterials.6.045004}, antisymmetric thermopolarization~\cite{PhysRevB.105.245125}, nonlinear transverse magnetization~\cite{JPSJ.92.043701}, and unconventional Hall effect~\cite{PhysRevB.108.085124}.

\begin{figure}[htbp]
  \centering
  \includegraphics[width=\linewidth]{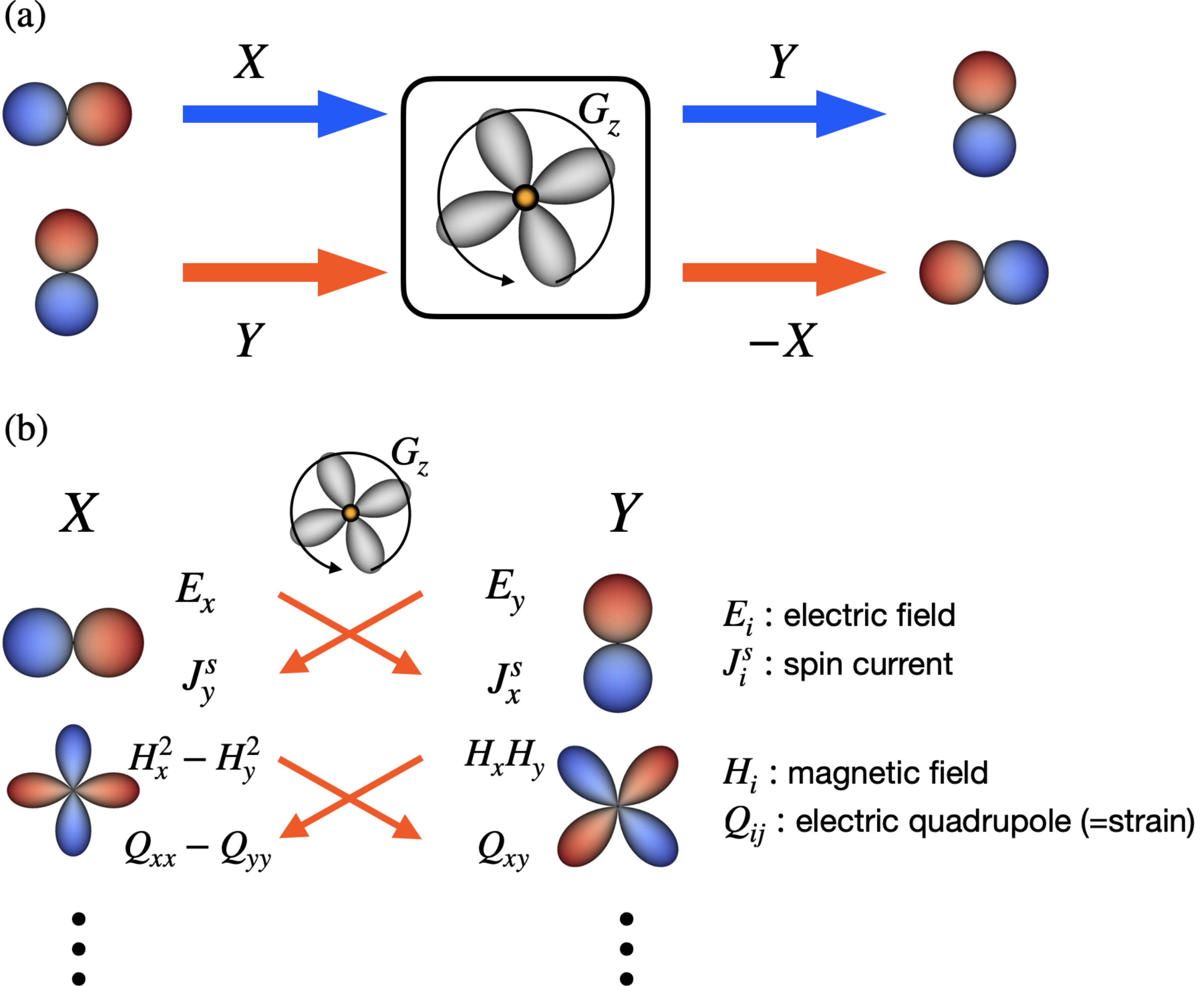}
  \caption{
    (a) Schematic picture of the rotational response under the electric toroidal (ET) dipole ($G_z$). 
    (b) The relation between the angle dependence of $X$ and $Y$, coupled by the ET dipole.
  }
  \label{Fig_1}
\end{figure}

The present study investigates such responses characteristic of ferroaxial ordering by focusing on its relation to a tensor transformation property under the symmetry operations.
The ET dipole under the ferroaxial ordering leads to a rotational response in the physical tensor with various ranks.
For example, when considering two different physical quantities $X$ and $Y$ with the same symmetry, the ET dipole gives rise to the rotational response for $X$ and $Y$; in the case of the rank-2 tensor, $Y$ ($-X$) is induced by applying $X$ ($Y$), as shown in Fig.~\ref{Fig_1}(a); a similar argument has been done in previous studies as mentioned above~\cite{JPSJ.91.113702, PhysRevB.105.245125, PhysRevB.108.085124}.
In other words, the sign change of responses occurs when input and output are interchanged, which means the rotational response.
By extending such a rotational response to higher-rank tensors, we find that a rotational tensor component of nonlinear magnetostriction, where the strain is induced by a second-order magnetic field, also occurs under the ferroaxial ordering, as schematically shown in Fig.~\ref{Fig_1}(b); the $x^{2}-y^{2}$(or $xy$)-type strain is induced by applying the magnetic field $H_{x}H_{y}$(or $H_{x}^{2}-H_{y}^{2}$).
We discuss its behavior by analyzing a fundamental $d$-orbital model under the tetragonal symmetry. 

{\it{Rotational response.}}---
First, we discuss the symmetry analysis of the response tensor.
We consider a general situation, where a physical quantity $B_{\mu}$ is induced by an external field $F_{\nu}$ ($\mu,\nu$ is the index for the component):
\begin{equation}
  B_{\mu}=\sum_{\nu}\chi_{\mu;\nu}F_{\nu}.
  \label{response_general}
\end{equation}
For example, according to Fig.~\ref{Fig_1}(b), $F_{\nu}$ corresponds to the electric field $E_{x}$ or the second-order magnetic field $H_{x}H_{y}$ and $B_{\mu}$ corresponds to the spin current $J_{x}^{s}$ or the strain $Q_{xx}-Q_{yy}$.
We decpomse $\chi_{\mu;\nu}$ into a symmetric part $\chi_{\mu;\nu}^{(\mathrm{S})}$ and an antisymmetric part $\chi_{\mu;\nu}^{(\mathrm{A})}$ as follows: 
\begin{equation}
  \chi_{\mu;\nu}=\chi_{\mu;\nu}^{(\mathrm{S})}+\chi_{\mu;\nu}^{(\mathrm{A})},
  \label{separtaion_tensor}
\end{equation}
where $\chi_{\mu;\nu}^{(\mathrm{S})}$ ($\chi_{\mu;\nu}^{(\mathrm{A})}$) is symmetric (antisymmetric) for the interchange of $(\mu\leftrightarrow\nu)$;
$\chi_{\mu;\nu}^{(\mathrm{S})}=\chi_{\nu;\mu}^{(\mathrm{S})}$ and $\chi_{\mu;\nu}^{(\mathrm{A})}=-\chi_{\nu;\mu}^{(\mathrm{A})}$.

As the transformation properties of $\chi_{\mu;\nu}^{(\mathrm{S})}$ and $\chi_{\mu;\nu}^{(\mathrm{A})}$ are different, their corresponding multipoles are different~\cite{PhysRevB.104.054412}.
To discuss nonlinear magnetostriction later, we consider the time-reversal-even rank-4 polar tensor $\chi^{[2\times 2]}$ by setting $B^{[2]}=(B_{xx}, B_{yy}, B_{zz}, B_{yz}, B_{zx}, B_{xy})$ and $F^{[2]}=(F_{xx}, F_{yy}, F_{zz}, F_{yz}, F_{zx}, F_{xy})$ with $B_{ij}=B_{ji}$ and $F_{ij}=F_{ji}$; the superscript of $B$ and $F$ represents the rank.
The relevant multipoles for $\chi^{[2\times 2]}$ are the rank-0--4: the rank-0 electric (E) monopole, the rank-1 ET dipole, the rank-2 E quadrupole, the rank-3 ET octupole, and the rank-4 E hexadecapole.
The ET dipole and octupole (E monopole and hexadecapole) appear only in $\chi^{(\mathrm{A})}(\chi^{(\mathrm{S})})$, while the E quadrupole appears in both tensor parts.
We show the relation between the tensor components and the multipoles in Table~\ref{component_MP}.
For example, let us focus on the tensor components $\chi_{xx;xy},\chi_{xy;xx}$.
From Table~\ref{component_MP}, one finds that these tensor components can become nonzero when the expectation values of the ET dipole $G_{z}$, E quadrupole $Q_{xy}$, ET octupoles $G_{z}^{\alpha}$ and $G_{z}^{\beta}$, and E hexadecapoles $Q_{4z}^{\alpha}$ and $Q_{4z}^{\beta}$ are finite.
Among them, $G_{z}$, $G_{z}^{\alpha}$, and $G_{z}^{\beta}$ contribute to the pure antisymmetric rotational response $\chi_{xx;xy}=-\chi_{xy;xx}$, $Q_{4z}^{\alpha}$ and $Q_{4z}^{\beta}$ contribute to the symmetric response $\chi_{xx;xy}=\chi_{xy;xx}$, and $Q_{xy}$ contribute to both responses.

\begin{table}[t]
  \caption{Relation between tensor components and 
  multipoles (MPs), where $Q_{0}$ is the electric (E) monopole, $(Q_{u},Q_{v},Q_{yz},Q_{zx},Q_{xy})$ is the E quadrupole, $(Q_{4},Q_{4u},Q_{4v},Q_{4x}^{\alpha},Q_{4y}^{\alpha},Q_{4z}^{\alpha},Q_{4x}^{\beta},Q_{4y}^{\beta},Q_{4z}^{\beta})$ is the E hexadecapole, $(G_{x},G_{y},G_{z})$ is the electric toroidal (ET) dipole, and $(G_{xyz},G_{x}^{\alpha},G_{y}^{\alpha},G_{z}^{\alpha},G_{x}^{\beta},G_{y}^{\beta},G_{z}^{\beta})$ is the ET octupole.
  For each E monopole and E quadrupole in the symmetric tensor, two of the three are independent in $\chi_{\mathrm{A}},\chi_{\mathrm{B}}$, and $\chi_{\mathrm{C}}$~\cite{comment}.}
  \centering
  \begin{tabular}{l|l}
    \hline\hline
    MP & antisymmetric $\chi_{\mu;\nu}^{(\mathrm{A})}=-\chi_{\nu;\mu}^{(\mathrm{A})}$
    \\
    \hline
    $G_{x}$ & $\chi_{yy;yz}=-\chi_{zz;yz}=2\chi_{xy;zx}=-\chi_{yz;yy}=\chi_{yz;zz}=-2\chi_{zx;xy}$ 
    \\
    $G_{y}$ & $\chi_{zz;zx}=-\chi_{xx;zx}=2\chi_{yz;xy}=-\chi_{zx;zz}=\chi_{zx;xx}=-2\chi_{xy;yz}$ 
    \\
    $G_{z}$ & $\chi_{xx;xy}=-\chi_{yy;xy}=2\chi_{zx;yz}=-\chi_{xy;xx}=\chi_{xy;yy}=-2\chi_{yz;zx}$ 
    \\
    \hline
    $Q_{u}$ & $\chi_{xx;zz}=\chi_{yy;zz}=-\chi_{zz;xx}=-\chi_{zz;yy}$ 
    \\
    $Q_{v}$ & $\chi_{xx;yy}=2\chi_{xx;zz}=-2\chi_{yy;zz}=-\chi_{yy;xx}=-2\chi_{zz;xx}=2\chi_{zz;yy}$ 
    \\
    $Q_{yz}$ & $\chi_{xx;yz}=\chi_{yy;yz}=\chi_{zz;yz}=-\chi_{yz;xx}=-\chi_{yz;yy}=-\chi_{yz;zz}$
    \\
    $Q_{zx}$ & $\chi_{xx;zx}=\chi_{yy;zx}=\chi_{zz;zx}=-\chi_{zx;xx}=-\chi_{zx;yy}=-\chi_{zx;zz}$
    \\
    $Q_{xy}$ & $\chi_{xx;xy}=\chi_{yy;xy}=\chi_{zz;xy}=-\chi_{xy;xx}=-\chi_{xy;yy}=-\chi_{xy;zz}$
    \\
    \hline
    $G_{xyz}$ & $\chi_{xx;yy}=\chi_{yy;zz}=\chi_{zz;xx}=-\chi_{yy;xx}=-\chi_{zz;yy}=-\chi_{xx;zz}$ 
    \\
    $G_{x}^{\alpha}$ & $2\chi_{yy;yz}=-2\chi_{zz;yz}=-\chi_{xy;zx}=-2\chi_{yz;yy}=2\chi_{yz;zz}=\chi_{zx;xy}$ 
    \\
    $G_{y}^{\alpha}$ & $2\chi_{zz;zx}=-2\chi_{xx;zx}=-\chi_{yz;xy}=-2\chi_{zx;zz}=2\chi_{zx;xx}=\chi_{xy;yz}$ 
    \\
    $G_{z}^{\alpha}$ & $2\chi_{xx;xy}=-2\chi_{yy;xy}=-\chi_{zx;yz}=-2\chi_{xy;xx}=2\chi_{xy;yy}=\chi_{yz;zx}$ 
    \\
    $G_{x}^{\beta}$ & $\chi_{xx;yz}=-2\chi_{yy;yz}=-2\chi_{zz;yz}=-\chi_{yz;xx}=2\chi_{yz;yy}=2\chi_{yz;zz}$ 
    \\
    $G_{y}^{\beta}$ & $\chi_{yy;zx}=-2\chi_{zz;zx}=-2\chi_{xx;zx}=-\chi_{zx;yy}=2\chi_{zx;zz}=2\chi_{zx;xx}$ 
    \\
    $G_{z}^{\beta}$ & $\chi_{zz;xy}=-2\chi_{xx;xy}=-2\chi_{yy;xy}=-\chi_{xy;zz}=2\chi_{xy;xx}=2\chi_{xy;yy}$ 
    \\
    \hline\hline
    MP & symmetric $\chi_{\mu;\nu}^{(\mathrm{S})}=\chi_{\nu;\mu}^{(\mathrm{S})}$
    \\
    \hline
    $Q_{0}$ & $\chi_{\mathrm{A}}:\chi_{xx;xx}=\chi_{yy;yy}=\chi_{zz;zz}$, $\chi_{\mathrm{B}}:\chi_{yz;yz}=\chi_{zx;zx}=\chi_{xy;xy}$ 
    \\
    & $\chi_{\mathrm{C}}:\chi_{xx;yy}=\chi_{yy;zz}=\chi_{zz;xx}=\chi_{yy;xx}=\chi_{zz;yy}=\chi_{xx;zz}$ 
    \\
    \hline
    $Q_{u}$ & $\chi_{\mathrm{A}}:2\chi_{xx;xx}=2\chi_{yy;yy}=-\chi_{zz;zz}$, $\chi_{\mathrm{B}}:2\chi_{yz;yz}=2\chi_{zx;zx}=-\chi_{xy;xy}$, 
    \\
    & $\chi_{\mathrm{C}}:2\chi_{xx;zz}=2\chi_{yy;zz}=-\chi_{xx;yy}=2\chi_{zz;xx}=2\chi_{zz;yy}=-\chi_{yy;xx}$ 
    \\
    $Q_{v}$ & $\chi_{\mathrm{A}}:\chi_{xx;xx}=-\chi_{yy;yy}$, $\chi_{\mathrm{B}}:\chi_{yz;yz}=-\chi_{zx;zx}$, 
    \\
    & $\chi_{\mathrm{C}}:\chi_{xx;zz}=-\chi_{yy;zz}=\chi_{zz;xx}=-\chi_{zz;yy}$ 
    \\
    $Q_{yz}$ & $\chi_{\mathrm{A}}:\chi_{xx;yz}=\chi_{yz;xx}$, $\chi_{\mathrm{B}}:\chi_{yy;yz}=\chi_{zz;yz}=\chi_{yz;yy}=\chi_{yz;zz}$, 
    \\
    & $\chi_{\mathrm{C}}:\chi_{zx;xy}=\chi_{xy;zx}$ 
    \\
    $Q_{zx}$ & $\chi_{\mathrm{A}}:\chi_{yy;zx}=\chi_{zx;yy}$, $\chi_{\mathrm{B}}:\chi_{zz;zx}=\chi_{xx;zx}=\chi_{zx;zz}=\chi_{zx;xx}$,
    \\
    & $\chi_{\mathrm{C}}:\chi_{xy;yz}=\chi_{yz;xy}$ 
    \\
    $Q_{xy}$ & $\chi_{\mathrm{A}}:\chi_{zz;xy}=\chi_{xy;zz}$, $\chi_{\mathrm{B}}:\chi_{xx;xy}=\chi_{yy;xy}=\chi_{xy;xx}=\chi_{xy;yy}$, 
    \\
    & $\chi_{\mathrm{C}}:\chi_{yz;zx}=\chi_{zx;yz}$ 
    \\
    \hline
    $Q_{4}$ & $\chi_{xx;xx}=\chi_{yy;yy}=\chi_{zz;zz}=-2\chi_{yz;yz}=-2\chi_{zx;zx}=-2\chi_{xy;xy}$
    \\
    $Q_{4u}$ & $2\chi_{xx;xx}=2\chi_{yy;yy}=-\chi_{zz;zz}=2\chi_{yz;yz}=2\chi_{zx;zx}=-\chi_{xy;xy}$
    \\
    $Q_{4v}$ & $\chi_{xx;xx}=-\chi_{yy;yy}=\chi_{yz;yz}=-\chi_{zx;zx}$
    \\
    $Q_{4x}^{\alpha}$ & $\chi_{yy;yz}=-\chi_{zz;yz}=\chi_{yz;yy}=-\chi_{yz;zz}$
    \\
    $Q_{4y}^{\alpha}$ & $\chi_{zz;zx}=-\chi_{xx;zx}=\chi_{zx;zz}=-\chi_{zx;xx}$
    \\
    $Q_{4z}^{\alpha}$ & $\chi_{xx;xy}=-\chi_{yy;xy}=\chi_{xy;xx}=-\chi_{xy;yy}$
    \\
    $Q_{4x}^{\beta}$ & $\chi_{xx;yz}=-2\chi_{yy;yz}=-2\chi_{zz;yz}=\chi_{zx;xy}$
    \\
    & $=\chi_{yz;xx}=-2\chi_{yz;yy}=-2\chi_{yz;zz}=\chi_{xy;zx}$
    \\
    $Q_{4y}^{\beta}$ & $\chi_{yy;zx}=-2\chi_{zz;zx}=-2\chi_{xx;zx}=\chi_{xy;yz}$
    \\
    & $=\chi_{zx;yy}=-2\chi_{zx;zz}=-2\chi_{zx;xx}=\chi_{yz;xy}$
    \\
    $Q_{4z}^{\beta}$ & $\chi_{zz;xy}=-2\chi_{xx;xy}=-2\chi_{yy;xy}=\chi_{yz;zx}$
    \\
    & $=\chi_{xy;zz}=-2\chi_{xy;xx}=-2\chi_{xy;yy}=\chi_{zx;yz}$
    \\
    \hline\hline
  \end{tabular}
  \label{component_MP}
\end{table}

The argument that the ET dipole only contributes to the antisymmetric component also holds for other even-rank polar tensors.
On the other hand, we note that the situation is different for the odd-rank axial tensor.
For example, let us consider the rank-3 tensor $\chi^{[1\times 2]}$ for $B^{[1]}=(B_{x},B_{y},B_{z})$ and $F^{[2]}=(F_{xx},F_{yy},F_{zz},F_{yz},F_{zx},F_{xy})$ with $F_{ij}=F_{ji}$.
In this case, the relevant multipoles are the rank-1--3: the rank-1 ET dipole, the rank-2 E quadrupole, and the rank-3 ET octupole, where the ET dipole can appear in both the symmetric and antisymmetric tensors, whereas the ET octupole (E quadrupole) appears only in the symmetric (antisymmetric) tensor; 
the detailed relation between tensor components and multipoles can be easily obtained by changing $(M, T) \to (Q, G)$ and $(\sigma^{\mathrm{O}}, \sigma^{\mathrm{H}}) \to (\chi^{(\mathrm{S})}, \chi^{(\mathrm{A})})$ in Eq.~(12) in Ref.~\citen{PhysRevB.107.155109}.
We summarize the relation between $\chi_{\mu;\nu}$ and the multipoles in Table~\ref{active_MP}.

\begin{table}[t]
  \caption{
  Relation between the parts of time-reversal-even rank-$3,4$ tensors $\chi^{(\mathrm{S})}/\chi^{(\mathrm{A})}$ and activated multipoles represented by \checkmark.
  }
  \centering
  \begin{tabular}{l|cc|cc}
    \hline\hline
    Multipole & rank-3 & (axial) & rank-4 & (polar) 
    \\
    & $\chi^{(\mathrm{S})}$ & $\chi^{(\mathrm{A})}$ & $\chi^{(\mathrm{S})}$ & $\chi^{(\mathrm{A})}$ 
    \\
    \hline
    E monopole & -- & -- & \checkmark & -- 
    \\
    ET dipole & \checkmark & \checkmark & -- & \checkmark 
    \\
    E quadrupole & -- & \checkmark & \checkmark & \checkmark 
    \\
    ET octupole & \checkmark & -- & -- & \checkmark 
    \\
    E hexadecapole & -- & -- & \checkmark & -- 
    \\
    \hline\hline
  \end{tabular}
  \label{active_MP}
\end{table}

{\it{Model calculation.}}---
We demonstrate the above symmetry argument by performing the model analysis; we consider the second-order nonlinear magnetostriction tensor, which is obtained by substituting $F^{[2]}=(H_{x}^{2}, H_{y}^{2}, H_{z}^{2}, H_{y}H_{z}, H_{z}H_{x}, H_{x}H_{y})$ and $B^{[2]}=(Q_{xx}, Q_{yy}, Q_{zz}, Q_{yz}, Q_{zx}, Q_{xy})$ in Eq.~(\ref{response_general}) as follows:
\begin{equation}
  Q_{\mu}=\sum_{ij}\chi_{\mu;ij}H_{i}H_{j}, 
  \label{quadrupole_H}
\end{equation}
where $Q_{\mu}$ is the rank-2 E quadrupole, which mimics the strain by the electronic origin; for example, $Q_{xx}$ corresponds to the $x^2$ component of the strain, i.e., $\varepsilon_{xx}$.
It is noted that one can obtain a qualitatively similar result for the second-order electric field $E_{i}E_{j}$ instead of $H_{i}H_{j}$, since they have the same symmetry, and the tensor is called as an electrostriction tensor~\cite{AdvPhys.3.85,ApplPhysRev.1.011103}.
Meanwhile, the elastic stiffness tensor with $\chi_{\mu;ij}=\chi_{ij;\mu}$ under $H_{i}H_{j}\to Q_{ij}$ is not appropriate; the contributions from the ET dipole and octupole vanish in this case.

The model is chosen to include the ET dipole degree of freedom in the Hilbert space; we specifically take five $d$-orbitals $(\phi_{u},\phi_{v},\phi_{yz},\phi_{zx},\phi_{xy})$ for $u=3z^{2}-r^{2},v=x^{2}-y^{2}$, with $d^{1}$ configuration under the point group $D_{\mathrm{4h}}$; the multipoles$Q_{0}$, $Q_{u}$, $Q_{4}$, and $Q_{4u}$ belong to the totally-symmetric irreducible representation.
The single-site Hamiltonian is given by 
\begin{equation}
  \mathcal{H}=\mathcal{H}_{\mathrm{loc}}+\mathcal{H}_{\mathrm{int}}.
\end{equation}
The first term represents the atomic Hamiltonian, which is represented by 
\begin{equation}
  \mathcal{H}_{\mathrm{loc}}=\sum_{\alpha\beta}h^{(0)}_{\alpha\beta}\hat{c}_{\alpha}^{\dag}\hat{c}_{\beta},
\end{equation}
where $\hat{c}_{\alpha}^{\dag}(\hat{c}_{\alpha})$ is the creation (annihilation) operator of an electron with internal degrees of freedom $\alpha$.
The Hamiltonian matrix $h^{(0)}_{\alpha\beta}=\braket{\alpha|\hat{h}^{(0)}|\beta}$ consists of two parts: 
\begin{equation}
  \hat{h}^{(0)}=\hat{h}_{\mathrm{CEF}}+\hat{h}_{\mathrm{SOC}}
\end{equation}
with 
\begin{subequations}
  \begin{equation}
    \hat{h}_{\mathrm{CEF}}=B_{20}Q_{u}+B_{40}Q_{40}+B_{44}Q_{44},
  \end{equation}
  \begin{equation}
    \hat{h}_{\mathrm{SOC}}=\lambda\bm{l}\cdot\bm{s}.
  \end{equation}
\end{subequations}
$\hat{h}_{\mathrm{CEF}}$ denotes the crystalline electric field (CEF) under $D_{\mathrm{4h}}$ and $Q_{40}, Q_{44}$ are spinless E hexadecapole operators, which are rewritten by a linear combination of $Q_{4}$ and $Q_{4u}$~\cite{PhysRevB.98.165110, JPSJ.87.033709}.
$\hat{h}_{\mathrm{SOC}}$ stands for the atomic spin--orbit coupling where $\bm{l}$ and $\bm{s}$ represent the orbital- and spin-angular momentum operators, respectively.
The interaction Hamiltonian $\mathcal{H}_{\mathrm{int}}$ is given by 
\begin{equation}
  \mathcal{H}_{\mathrm{int}}=-J\hat{G}_{z}\hat{G}_{z},
\end{equation}
where we consider an effective interaction favoring the uniform ET dipole (ferroaxial) ordering; $J>0$ is the coupling constant.
$\hat{G}_{z}$ is defined by 
\begin{equation}
  \hat{G}_{z}\equiv \sum_{\alpha\beta}(G_{z})_{\alpha\beta}\hat{c}_{\alpha}^{\dag}\hat{c}_{\beta}
\end{equation}
with~\cite{JPSJ.91.113702,PhysRevLett.119.187203,PhysRevB.104.235125,PhysRevLett.130.256801} 
\begin{equation}
  G_{z}=\frac{1}{\sqrt{5}}(\bm{l}\times\bm{s})_{z},
\end{equation}
where $G_{z}$ is normalized to satisfy $\mathrm{Tr}[G_{z}^{2}]=2$.
We apply the mean-field approximation for this term as 
\begin{equation}
  \hat{\mathcal{H}}_{\mathrm{int}}\simeq\sum_{\alpha\beta}\left(h_{\mathrm{int}}^{\mathrm{MF}}\right)_{\alpha\beta}\hat{c}_{\alpha}^{\dag}\hat{c}_{\beta}+\mathrm{const.},
\end{equation}
where $\hat{h}_{\mathrm{int}}^{\mathrm{MF}}=-J\braket{\hat{G}_{z}}G_{z}$ with a thermal average $\braket{\cdots}$.
In the following, we choose $J$ as the unit of energy of this model.
We choose the other parameters as $B_{20}=0.1J,B_{40}=0.3J,B_{44}=0.2J$, and $\lambda=0.1J$.

\begin{figure}[htbp]
  \centering
  \includegraphics[width=\linewidth]{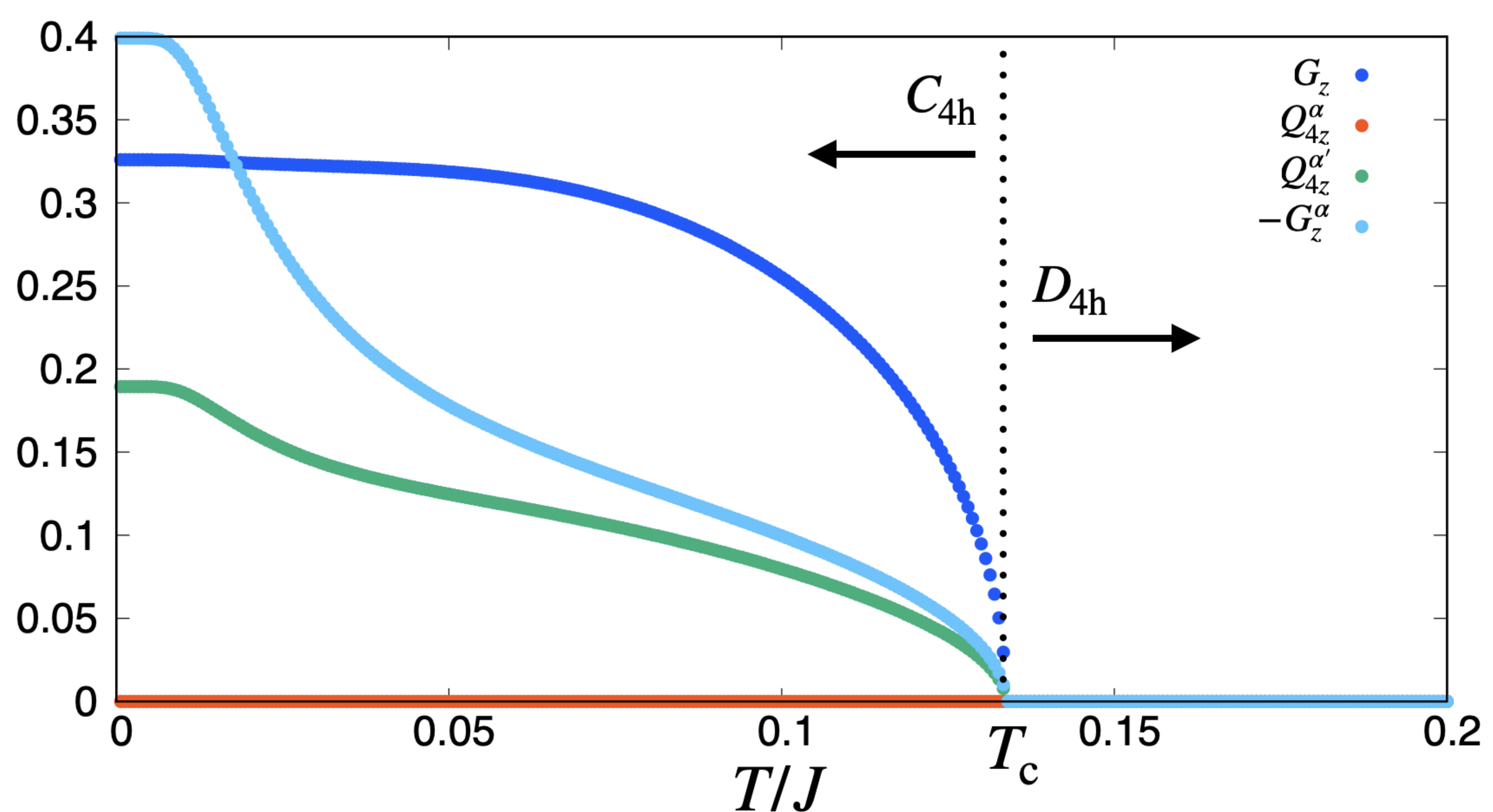}
  \caption{
    The temperature dependence of $A_{2g}^{+}$ multipoles (order parameters) by the ET dipole--ET dipole interaction within the mean-field approximation.
    All multipole operators $X$ are normalized to satisfy $\mathrm{Tr}[X^{2}]=2$.
    The vertical dashed line represents the phase transition temperature $T_{\mathrm{c}}$.
  }
  \label{Fig_2}
\end{figure}

By performing the self-consistent mean-field calculations for $\hat{h}^{(0)}+\hat{h}_{\mathrm{int}}^{\mathrm{MF}}$, we obtain the phase transition to ferroaxial ordering at finite temperatures as shown in Fig.~\ref{Fig_2}.
The phase transition temperature is given by $T_{\mathrm{c}}\simeq 0.135 J$, and the symmetry of the system is lowered from $D_{\mathrm{4h}}$ to $C_{\mathrm{4h}}$ for $T<T_{\mathrm{c}}$; $G_{z}$ belongs to the totally-symmetric irreducible representation. 
Note that the present model includes other three multipole degrees of freedom with the same symmetry as $G_{z}$; ET octupole $G_{z}^{\alpha}$, spinless E hexadecapole $Q_{4z}^{\alpha}\propto xy(x^{2}-y^{2})$, and spinful E hexadecapole $Q_{4z}^{\alpha\prime}$.
As shown in Fig.~\ref{Fig_2}, $Q_{4z}^{\alpha\prime}$ and $G_{z}^{\alpha}$ are also induced, while $Q_{4z}^{\alpha}$ is not.
Here, $Q_{4z}^{\alpha\prime}$ and $G_{z}^{\alpha}$ are given by
\begin{equation*}
  \begin{aligned}
    G_{z}^{\alpha}\propto&\, M_{3u}s_{y}-M_{3v}s_{x},
    \\
    Q_{4z}^{\alpha\prime}\propto&\, M_{3a}s_{y}+M_{3b}s_{x},
    \\
  \end{aligned}
\end{equation*}
where $M_{3u}$, $M_{3v}$, $M_{3a}$, and $M_{3b}$ are magnetic octupole operators~\cite{JPSJ.89.104704}.
Among three nonzero multipoles, $Q_{4z}^{\alpha\prime}$ vanishes for $B_{44}=0$, while the others do not, which indicates that the conditions to activate E and ET multipoles in terms of the model parameters are different.

We investigate the second-order nonlinear magnetostriction in Eq.~(\ref{quadrupole_H}).
Here, we evaluate the expectation value of E quadrupoles induced by the magnetic field.
For that purpose, we introduce the Zeeman Hamiltonian, which is given by 
\begin{equation}
  \hat{h}_{\mathrm{Zeeman}}=-\mu_{\mathrm{B}}\bm{H}\cdot(\bm{l}+2\bm{s})
\end{equation}
with $\bm{H}=(H_{x},H_{y},H_{z})$.
We set the Bohr magneton as unity, i.e., $\mu_{\mathrm{B}}=1$. 

First, we discuss nonzero components of the tensor $\chi_{\mu;ij}$ in Eq.~(\ref{quadrupole_H}) under the ferroaxial ordering from the symmetry.
Under the $D_{\mathrm{4h}}$ symmetry, nonzero tensor components are as follows: 
$\chi_{xx;xx}^{(\mathrm{S})}=\chi_{yy;yy}^{(\mathrm{S})},\chi_{zz;zz}^{(\mathrm{S})}$, $\chi_{yz;yz}^{(\mathrm{S})}=\chi_{zx;zx}^{(\mathrm{S})},\chi_{xy;xy}^{(\mathrm{S})}$, $\chi_{xx;yy}^{(\mathrm{S})}, \chi_{yy;zz}^{(\mathrm{S})}=\chi_{zz;xx}^{(\mathrm{S})}$, and $\chi_{xx;zz}^{(\mathrm{A})}=\chi_{yy;zz}^{(\mathrm{A})}$, which reflects the activation of $Q_0$, $Q_u$, $Q_4$, and $Q_{4u}$, as described above.
In addition, the following tensor components can become nonzero once the ferroaxial ordering occurs: 
$\chi_{xx;xy}^{(\mathrm{S})}=-\chi_{yy;xy}^{(\mathrm{S})}$, $\chi_{xx;xy}^{(\mathrm{A})}=-\chi_{yy;xy}^{(\mathrm{A})}$, and $\chi_{yz;zx}^{(\mathrm{A})}$ owing to the activation of $G_{z}$, $G^{\alpha}_{z}$, and $Q^{\alpha\prime}_{4z}$.
Based on Table~\ref{component_MP}, $\chi_{xx;xy}^{(\mathrm{S})}$ is related to the E hexadecapole $Q_{4z}^{\alpha}$ or $Q_{4z}^{\alpha\prime}$, whereas $\chi_{xx;xy}^{(\mathrm{A})}$ and $\chi_{yz;zx}^{(\mathrm{A})}$ are related to the ET dipole $G_{z}$ and/or ET octupole $G_{z}^{\alpha}$.
By appropriately taking the magnetic-field direction, the correspondence between the E quadrupole and the tensor component under the $C_{\mathrm{4h}}$ symmetry is given by 
\begin{equation}
  \begin{aligned}
    \chi_{xx;xy}^{(\mathrm{S})}=&\, \frac{Q_{v}(\pi/2,\pi/4)+Q_{xy}(\pi/2,0)}{H^{2}},
    \\
    \chi_{xx;xy}^{(\mathrm{A})}=&\, \frac{Q_{v}(\pi/2,\pi/4)-Q_{xy}(\pi/2,0)}{H^{2}},
    \\
    \chi_{yz;zx}^{(\mathrm{A})}=&\, \frac{Q_{yz}(\pi/4,0)}{H^{2}}=-\frac{Q_{zx}(\pi/4,0)}{H^{2}}, 
    \\
  \end{aligned}
  \label{E_quadrupole}
\end{equation}
where $Q_{\mu}(\theta,\varphi)$ represents the E quadrupoles as a function of the magnetic-field direction with $\bm{H}=H(\cos{\varphi}\sin{\theta},\sin{\varphi}\sin{\theta},\cos{\theta});0\leq \theta <\pi, 0\leq \varphi <2\pi$.
Thus, one can distinguish the contributions from E multipoles and ET multipoles by changing the magnetic-field direction and stress component.

\begin{figure}[htbp]
  \centering
  \includegraphics[width=\linewidth]{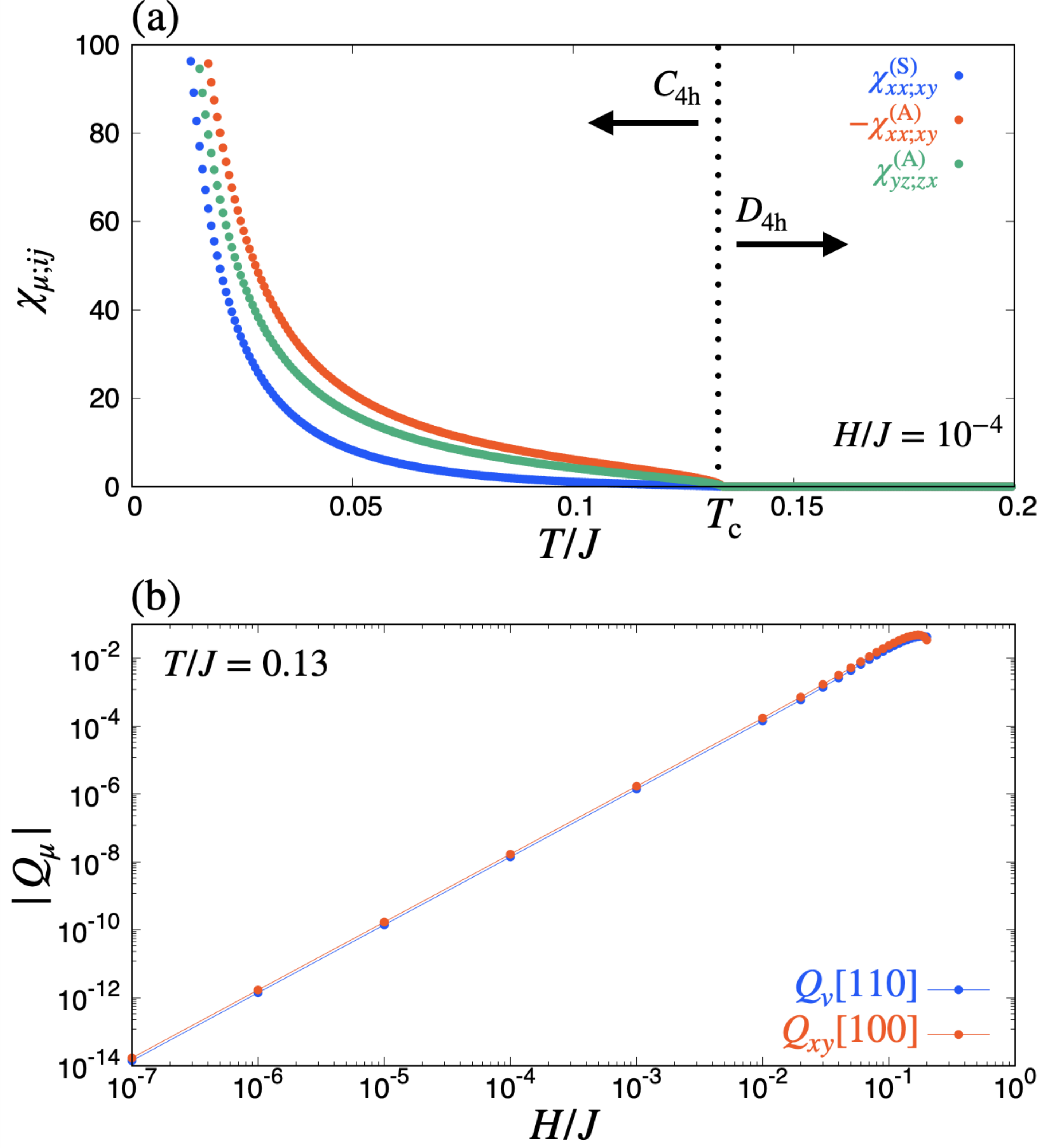}
  \caption{
    (a) The temperature dependence of the second-order magnetostriction response tensor induced by the ET dipole ordering at magnetic field $H=10^{-4}J$.
    The vertical dashed line represents the phase transition temperature $T_{\mathrm{c}}$.
    (b) The magnetic field strength dependence of expectation values of electric quadrupoles at $T=0.13J$ (ordered phase).
    $[110]$ and $[100]$ mean the direction of the magnetic field.
  }
  \label{Fig_3}
\end{figure}

Figure~\ref{Fig_3}(a) shows the temperature dependence of second-order magnetostriction tensors in Eq.~(\ref{E_quadrupole}).
The above additional components are nonzero below $T_{\mathrm{c}}$.
The nonzero $\chi^{(\mathrm{A})}_{xx;xy}$ and $\chi_{yz;zx}^{(\mathrm{A})}$ indicate that $\braket{G_{z}}$ and/or $\braket{G_{z}^{\alpha}}$ contribute to the rotational response.
On the other hand, the $\chi_{xx;xy}^{(\mathrm{S})}$ shows the contribution of $\braket{Q_{4z}^{\alpha\prime}}$.
In addition, we confirm that $\chi_{xx;xy}^{(\mathrm{S})}$ vanishes by setting $B_{44}=0$.
We also confirm such behaviors by calculating the essential model parameter~\cite{JPSJ.91.014701} of these tensors; we find that $\chi_{xx;xy}^{(\mathrm{A})}$ and $\chi_{yz;zx}^{(\mathrm{A})}$ are proportional to $\braket{\hat{G}_{z}}$, whereas $\chi_{xx;xy}^{(\mathrm{S})}$ is proportional to $B_{44}\braket{\hat{G}_{z}}$.
Then, we show the magnetic field strength dependence of E quadrupoles $Q_{v}(\pi/2,\pi/4)$ and $Q_{xy}(\pi/2,0)$ in Fig.~\ref{Fig_3}(b).
One finds that the E quadrupoles are proportional to $H^{2}$ in the low-$H$ region up to about $10^{-1}J$, which indicates that the effect of higher-order magnetic fields is negligible in Eq.~(\ref{quadrupole_H}) for small $H$.

\begin{figure}[htbp]
  \centering
  \includegraphics[width=\linewidth]{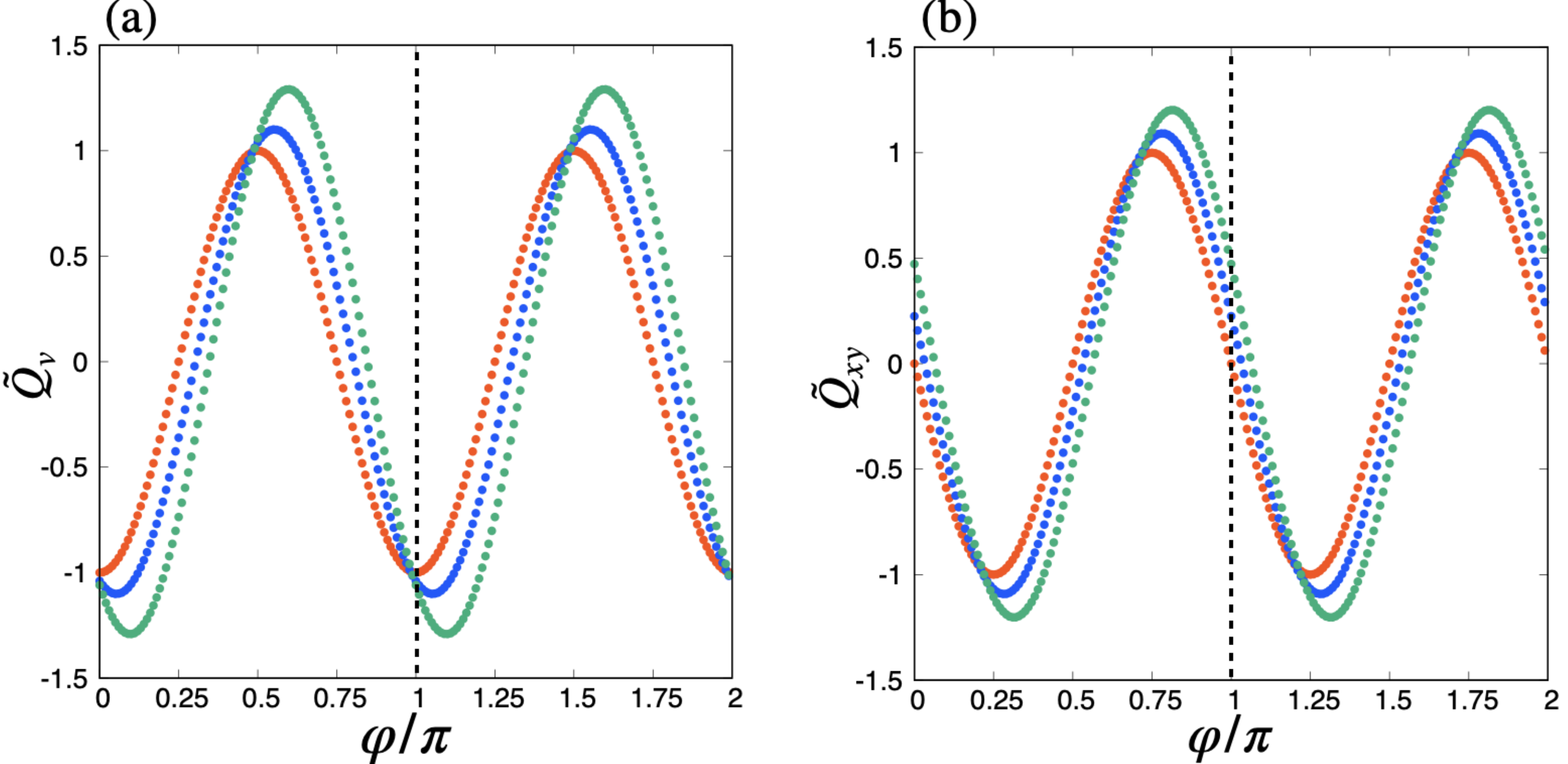}
  \caption{
    In-plane magnetic field $(\theta=\pi/2)$ dependence of electric quadrupoles with $H=10^{-1}J$.
    The red dots result at $T=0.14J$ in the para phase, whereas the blue (green) dots result at $T=0.13J$ ($T=0.12J$) in the ferroaxial phase.
    $\tilde{Q}_{\mu}$ denotes the normalized results by the maximum value of $T=0.14J$.
    The black dashed line indicates the angle at $\varphi=\pi$.
  }
  \label{Fig_4}
\end{figure}

Finally, we discuss how to detect the rotational response in experiments.
Figure~\ref{Fig_4} shows the angle dependence of E quadrupoles $Q_{v}, Q_{xy}$ above and below $T_{\mathrm{c}}$.
One finds that $x^{2}-y^{2}$($xy$)-type distortion are maximum at $\varphi=\pi/2$ and $3\pi/2$ ($\varphi=3\pi/4$ and $7\pi/4$) in the para phase; this feature is independent of the temperature.
When the ferroaxial ordering occurs, the angle of the maximum strain is shifted and depends on temperature.
Therefore, the rotational response can be detected by measuring the strain while changing the direction of the magnetic field and/or the temperature.

{\it{Conclusion.}}---
We theoretically investigated the rotational response by the ET dipole based on symmetry and microscopic model analyses.
We clarified that the ordering of the ET dipole and octupole induce the antisymmetric response, whereas the E hexadecapole induces the symmetric response.
We also discussed the nonlinear magnetostriction for the five $d$-orbital model in the tetragonal system.
When the ferroaxial ordering occurs, the angle of the magnetic field dependence on strain shows characteristic, which can be detected in experiments.
Recently, the high-resolution magnetostriction measurement has been done to detect the FFLO state in $\mathrm{CeCoIn}_{5}$~\cite{PhysRevB.107.L220505}.
We expect that such a measurement will enable us to investigate the rotational responses driven by the ferroaxial ordering.
Although we have analyzed a specific model and symmetry to demonstrate nonlinear magnetostriction, our result can be straightforwardly applied to other models under different lattice structures, where the crystallographic point groups without vertical mirror symmetry are $C_{\mathrm{6h}}, C_{6}, C_{\mathrm{3h}}, C_{\mathrm{4h}}, C_{4}, S_{4}, C_{\mathrm{3i}}, C_{3}, C_{\mathrm{2h}}, C_{2}, C_{\mathrm{s}}, C_{\mathrm{i}}$, and $C_{1}$.

\acknowledgment
This research was supported by JSPS KAKENHI Grants Numbers JP21H01037, JP22H04468, JP22H00101, JP22H01183, JP23K03288, JP23H04869, and by JST PRESTO (JPMJPR20L8).

\bibliographystyle{jpsj.bst}
\bibliography{main_text.bbl}

\begin{thebibliography}{10}
\bibitem{comment}
For example, if we choose $\chi_{\mathrm{A}}$ and $\chi_{\mathrm{B}}$ as indepndent components, $\chi_{\mathrm{C}}$ is expressed as $\chi_{\mathrm{C}}=(\chi_{{\mathrm{A}}}-\chi_{\mathrm{B}})/2$. See Ref.~\citen{PhysRevB.104.054412} for the details of the independent components.
\end{thebibliography}

\end{document}